\documentclass[journal]{IEEEtran}
\usepackage{array,multirow}
\usepackage{geometry}
\usepackage{nicefrac}
\usepackage{graphicx}
\usepackage{amsmath}
\usepackage{amssymb}
\usepackage{setspace}
\usepackage{enumerate}
\usepackage{array}
\usepackage[ruled,linesnumbered,vlined]{algorithm2e}
\usepackage{bm}
\usepackage{cite}

\usepackage{microtype}
\usepackage{hyperref}

\usepackage{hyperref,color}

\definecolor{webgreen}{rgb}{0,.35,0}
\definecolor{webbrown}{rgb}{.6,0,0}
\definecolor{RoyalBlue}{rgb}{0,0,0.9}
\definecolor{purp}{rgb}{0.6,0.05,0.8}
\definecolor{ora}{rgb}{0.7,0.35,0.02}

\hypersetup{
   colorlinks=true, linktocpage=true, pdfstartpage=3, pdfstartview=FitV,
   breaklinks=true, pdfpagemode=UseNone, pageanchor=true, pdfpagemode=UseOutlines,
   plainpages=false, bookmarksnumbered, bookmarksopen=true, bookmarksopenlevel=1,
   hypertexnames=true, pdfhighlight=/O,
   urlcolor=webbrown, linkcolor=RoyalBlue, citecolor=webgreen,
   pdfauthor={Gary P. T. Choi, Bernard Chiu, Chris H. Rycroft},
   pdfsubject={Area-preserving mapping of 3D ultrasound carotid artery images using density-equalizing reference map},
   pdfkeywords={},
   pdfcreator={pdfLaTeX},
   pdfproducer={LaTeX with hyperref}
}

\newcolumntype{C}[1]{>{\centering\let\newline\\\arraybackslash\hspace{0pt}}m{#1}}

\newcommand{\p}{\partial}

\graphicspath{{figs/}}
\begin{document}

\title{Area-preserving mapping of 3D ultrasound carotid artery images using density-equalizing reference map}

\author{Gary P. T. Choi, Bernard Chiu, and Chris H. Rycroft\\
\thanks{This work was supported in part by the Croucher Foundation (to G. P. T. Choi), the Research Grant Council of HKSAR (Project Nos. CityU 11205917, CityU 11203218) (to B. Chiu), the City University of Hong Kong Strategic Research Grant (No. 7004617) (to B. Chiu), and the Applied Mathematics Program of the U.S. Department of Energy (DOE) Office of Advanced Scientific Computing Research under contract DE-AC02-05CH11231 (to C. H. Rycroft).}
\thanks{G. P. T. Choi is with the John A. Paulson School of Engineering and Applied Sciences, Harvard University, Cambridge, MA 02138, USA (email: pchoi@g.harvard.edu).}
\thanks{B. Chiu is with the Department of Electronic Engineering, City University of Hong Kong, Hong Kong (email: bcychiu@cityu.edu.hk).}
\thanks{C. H. Rycroft is with the John A. Paulson School of Engineering and Applied Sciences, Harvard University, Cambridge, MA 02138, USA, and the Mathematics Group, Lawrence Berkeley National Laboratory, Berkeley, CA 94720, USA (email: chr@seas.harvard.edu).}

}


\maketitle

\begin{abstract}
Carotid atherosclerosis is a focal disease at the bifurcations of the carotid artery. To quantitatively monitor the local changes in the vessel-wall-plus-plaque thickness (VWT) and compare the VWT distributions for different patients or for the same patients at different ultrasound scanning sessions, a mapping technique is required to adjust for the geometric variability of different carotid artery models. In this work, we propose a novel method called \emph{density-equalizing reference map} (DERM) for mapping 3D carotid surfaces to a standardized 2D carotid template, with an emphasis on preserving the local geometry of the carotid surface by minimizing the local area distortion. The initial map was generated by a previously described arc-length scaling (ALS) mapping method, which projects a 3D carotid surface onto a 2D non-convex L-shaped domain. A smooth and area-preserving flattened map was subsequently constructed by deforming the ALS map using the proposed algorithm that combines the density-equalizing map and the reference map techniques. This combination allows, for the first time, one-to-one mapping from a 3D surface to a standardized non-convex planar domain in an area-preserving manner. Evaluations using 20 carotid surface models show that the proposed method reduced the area distortion of the flattening maps by over 80\% as compared to the ALS mapping method. 
\end{abstract}


\IEEEpeerreviewmaketitle

\section{Introduction}
\IEEEPARstart{S}{troke} is a leading cause of death and disability worldwide, causing an annual mortality of nearly 133,000 in United States \cite{Benjamin17} and over 1.6 million in China \cite{Liu11}. Carotid atherosclerosis is a major source of emboli, which may travel and block one of the cerebral arteries, causing ischemic stroke \cite{Eicke95}. As atherosclerosis is a focal disease with plaques predominantly occurring at bifurcations (BFs) of the carotid artery, monitoring local changes in the vessel-wall-plus-plaque thickness (VWT) as introduced by Chiu \textit{et al.}~\cite{Chiu08a} is important for the development of sensitive biomarkers that can identify high-risk patients with rapid plaque progression in a shorter time frame. Although the VWT-Change distribution for an individual patient can be visualized by mapping point-by-point measurements onto the three-dimensional carotid vessel wall, a flattened representation facilitates image analysis as clinicians can scan the entire image in a systematic manner without having to rotate and interact with the 3D surface. 

Projection-based visualization has been widely studied for medical applications, such as colon unfolding \cite{Haker00,Bartroli01,Zeng10}, brain flattening \cite{Angenent99,Gu04,Wang07,Choi15}, and the visualization of bone \cite{Kok10} and myocardium \cite{Termeer08}. A more detailed discussion on existing projection-based medical visualization methods can be found in the recent surveys \cite{Kreiser18,Grossmann18}. Unless a surface has a uniformly zero Gaussian curvature, distortion in either angle or area comes with surface flattening and algorithms have been introduced to preserve angle (conformal) or area \cite{Floater05}. An area-preserving flattening algorithm is preferred because the size of the plaque burden can be quantified in an area-preserving flattened map and this is important for risk stratification \cite{Wannarong13} and evaluating treatment strategies \cite{Ainsworth05}. Two area-preserving carotid flattening methods have been previously introduced \cite{Zhu05,Chiu08}.

\begin{figure}[t]
\centering
\includegraphics[width=0.45\textwidth]{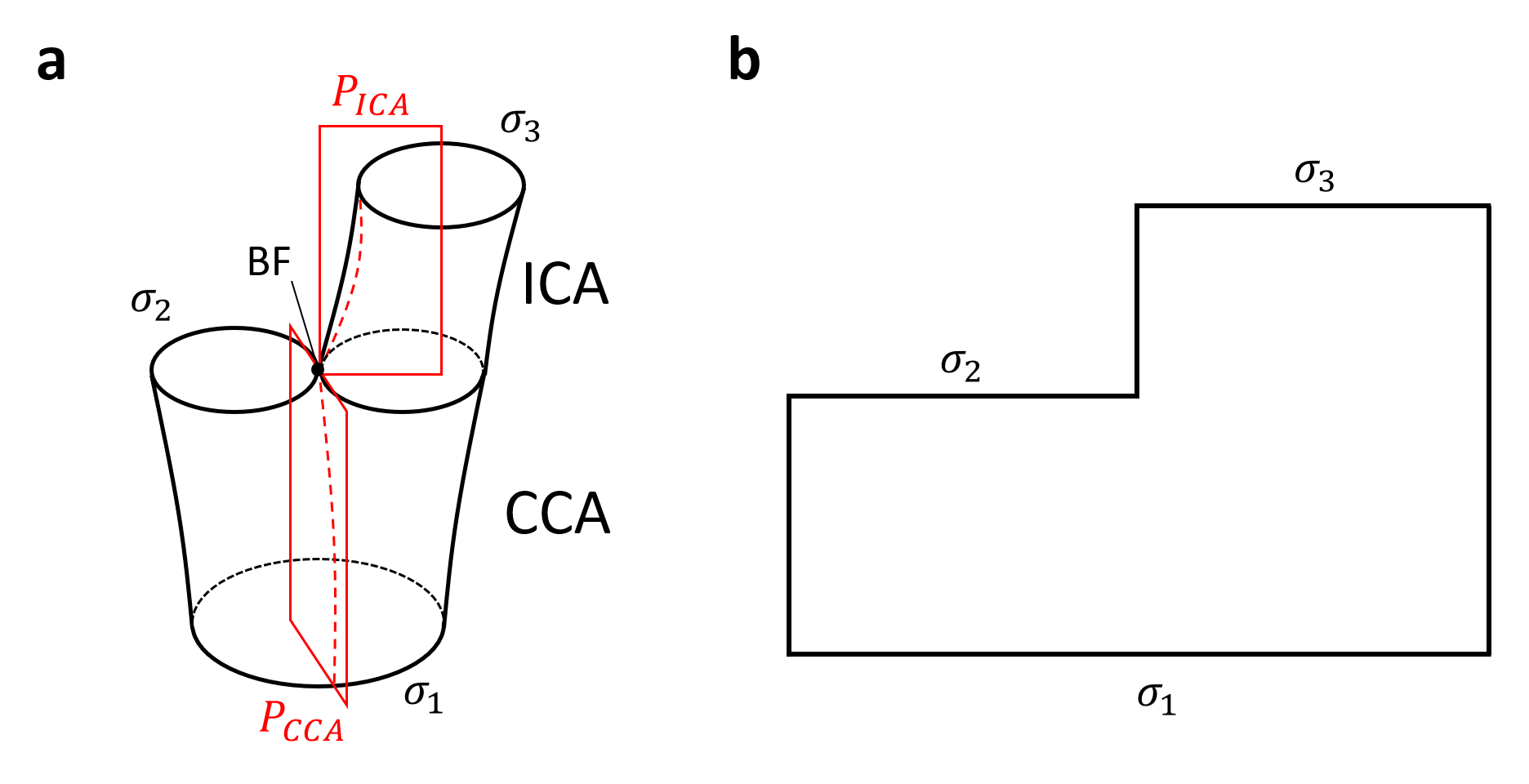}
\caption{An illustration of the 2D arc-length scaling (ALS) map \cite{Chiu13} of carotid surfaces. A carotid surface is first cut by two planes, denoted by $P_{\textit{ICA}}$ and $P_{\textit{CCA}}$ (as shown in \textbf{a}), and then unfolded to a 2D L-shaped non-convex domain (as shown in \textbf{b}). The arc-length of transverse contours segmented from 2D transverse images resliced from 3D ultrasound images is rescaled such that all vertices on the carotid surface correspond to uniformly spaced grid points on the L-shaped domain. The bifurcation (denoted by BF) is mapped to the non-convex corner of the L-shaped domain, and the carotid boundaries $\sigma_1, \sigma_2, \sigma_3$ are mapped to the horizontal boundaries of the L-shaped domain.}
\label{fig:carotid_artery_flattening}
\end{figure}

Zhu \textit{et al.} \cite{Zhu05} computed a conformal map projecting a 3D carotid surface onto a 2D plane first and then applied the optimal mass transport theory to correct for the area distortion exhibited in the conformal map. Chiu \textit{et al.} \cite{Chiu08} mapped each transverse cross-section of the carotid surface to a plane in an arc-length preserving manner and then solved the Monge-Kantorovich problem to correct for the area distortion occurred in the arc-length preserving map. This algorithm has been applied to multiple clinical studies \cite{Egger08,Krasinski09}. A major disadvantage of these two algorithms is that the shapes of the flattened maps are highly subject-specific and do not allow for a systematic and quantitative comparison between the VWT distribution of subjects who underwent different therapies or of the same subject derived from different imaging modalities. To address this issue, Chiu \textit{et al.} \cite{Chiu13} developed the arc-length scaling (ALS) mapping approach for mapping the 3D carotid surfaces onto a standardized 2D non-convex L-shaped domain (Fig.~\ref{fig:carotid_artery_flattening}). Note that the external carotid artery (ECA) (i.e. the left branch above the bifurcation point) is excluded in the analysis since plaques at ECA are not driectly related to stroke. The availability of the 2D standardized L-shaped carotid map has allowed for the development of sensitive biomarkers that are able to identify the effect of atorvastatin \cite{Chiu13,Chiu16} and B Vitamins \cite{Cheng17}. However, the ALS mapping approach does not consider any local geometric distortion introduced by the 3D to 2D mapping. Although methods have been introduced for improving the L-shaped flattening map by minimizing the descriptive length \cite{Chen16} and the angular distortion \cite{Choi17}, no area-preserving approach has been introduced for carotid flattening onto the standardized L-shaped domain. 

Recently, Choi and Rycroft \cite{Choi18} developed a method for computing surface flattening maps by extending a technique for cartogram projection called the density-equalizing maps \cite{Gastner04}. Although the shape-prescribed density-equalizing map method \cite{Choi18} is successful in producing an area-preserving map from a 3D surface onto a 2D domain, the bijectivity of the map is not guaranteed if the target 2D domain is non-convex. In this work, we developed and validated a novel method called \emph{density-equalizing reference map} (abbreviated as DERM) for computing area-preserving flattening maps of carotid surfaces onto the 2D non-convex L-shaped domain, using an improved formulation of density-equalizing maps. Our method extends the idea of density diffusion for handling the non-convex corner of the L-shaped domain, and further combines it with the reference map technique \cite{Kamrin12,Valkov15}, coupling the deformation of individual nodes with the spatial information for producing a smooth and accurate mapping result. To the best of our knowledge, this is the first work on producing a bijective area-preserving map from 3D surfaces onto a standardized non-convex 2D domain. The area-preserving map generated using the proposed algorithm will facilitate unbiased quantitative comparisons of the extent of carotid diseases among patients involved in population studies.  

\section{Mathematical background} \label{sect:background}
\subsection{Density-equalizing map} \label{sect:density_equalizing_map}
Gastner and Newman \cite{Gastner04} proposed an algorithm for producing \emph{density-equalizing map projections} based on a physical principle of density diffusion. Conceptually, given a planar map (such as the world atlas) and a density distribution prescribed on every part of the map, the method continuously deforms the map such that the difference in the density at different regions is transformed into a difference in the area of the regions. Regions with a larger prescribed density expand and those with a smaller density shrink. Ultimately, the density is equalized over the entire deformed map. 

Mathematically, given a planar domain $D$ and a quantity $\rho({\bf x},0)$ called the \emph{density} defined at every location ${\bf x}$ on $D$ at time $t = 0$, the method deforms $D$ by equalizing $\rho({\bf x},t)$ as $t \to \infty$. The equalization of $\rho$ follows from the advection equation
\begin{equation}
\frac{\partial \rho}{\partial t} = -\nabla \cdot {\bf j}, 
\end{equation}
where ${\bf j} = - \nabla \rho$ is the diffusion flux with uniform diffusivity. This yields the diffusion equation
\begin{equation}\label{eqt:diffusion}
\frac{\partial \rho}{\partial t} = \Delta \rho. 
\end{equation}
The density flux causes any tracers to move with velocity
\begin{equation}\label{eqt:velocity_field}
{\bf v} = - \frac{\nabla \rho}{\rho}.
\end{equation}
With the velocity field, the location of any tracer ${\bf x}$ at time $t$ can be traced by
\begin{equation}\label{eqt:displacement}
{\bf x}(t) = {\bf x}(0) + \int_0^t {\bf v}({\bf x}(\tau),\tau) d \tau.
\end{equation}
Taking $t \to \infty$, $\rho$ is equalized on the entire domain, and the above displacement produces a deformed map ${\bf x}_{\text{final}}$ that encodes all the density difference in the original density $\rho({\bf x},0)$ in terms of the area difference of different parts on the final deformed map ${\bf x}_{\text{final}}$. 

The algorithm was primarily applied to the visualization of sociological data such as population and average income on the world atlas \cite{Dorling10}. Recently, Choi and Rycroft \cite{Choi18} explored the close connection between density-equalizing maps and surface flattening maps. In particular, by setting $\rho$ as the face area of a 3D surface mesh and computing a density-equalizing map from the 3D surface to a 2D domain, the deformation follows the ratio of the face area at different regions and hence the resulting density-equalizing map is an area-preserving map. This idea sets the stage for our area-preserving mapping method for carotid surfaces.

\subsection{The reference map technique} \label{sect:reference_map}
Rycroft, Kamrin, and colleagues \cite{Kamrin12,Valkov15} developed the \emph{reference map technique} for simulating large-strain solid mechanics with a fully Eulerian formulation. Suppose a body is under deformation such that the material initially located at the position ${\bf X}$ is moved to the position ${\bf x}$ at time $t$. ${\bf x}({\bf X},t)$ is a motion function which keeps track of the motion of the material initially at ${\bf X}$. Define the \emph{reference map} ${\bf X} = \bm{\xi}({\bf x},t)$ as the inverse of the motion function. $\bm{\xi}({\bf x},t)$ can be regarded as a vector field in the deformed body which indicates the reference location of the material occupying the position ${\bf x}$ at time $t$. In particular, $\bm{\xi}({\bf x},0) = {\bf x}$. Since the reference location of any tracer particle is the same at all time $t$ under the deformation, we have $\dot{\bm{\xi}} = 0$, yielding the advection equation
\begin{equation} \label{eqt:reference_map}
\frac{\p \bm{\xi}}{\p t} + {\bf v} \cdot \nabla {\bm{\xi}} = {\bf 0}.
\end{equation}
By solving the above equation as $t \to \infty$, the final reference map $\bm{\xi}({\bf x}_{\text{final}}, \infty)$ gives the reference location of the material occupying the final position ${\bf x}_{\text{final}}$ of the deformed body. Therefore, to obtain ${\bf x}_{\text{final}}$ from ${\bm{\xi}}$, we simply need to track the contour lines of the $x$- and $y$- coordinates of ${\bm{\xi}}$.

\section{Materials and methods} \label{sect:main}
\subsection{Study subjects and 3D ultrasound image acquisition}
Ten subjects with carotid atherosclerosis were involved in this study. These subjects were recruited from The Premature Atherosclerosis Clinic at University Hospital (London Health Science Center, London, Canada) and the Stroke Prevention \& Atherosclerosis Research Center, Robarts Research Institute (London, Canada). Each subject was scanned at baseline and two weeks later with a 3D ultrasound carotid imaging system \cite{Fenster00}, providing 20 carotid ultrasound images for the evaluation of the proposed algorithm. 

The reconstruction of the 3D carotid surfaces and the construction of the 3D VWT map was described in the previous works by Chiu \textit{et al.} \cite{Chiu08a,Chiu13}. The 3D images were resliced at $1\text{~mm}$ intervals perpendicular to the medial axis of the vessel identified by an expert observer, who then segmented the outer wall and the lumen of the carotid artery on each resliced image. The 3D lumen and vessel wall surfaces were subsequently reconstructed from the segmented boundary stacks. To reduce the effect of segmentation variability, the lumen and outer wall boundaries were repeatedly segmented for five times, with consecutive segmentation sessions separated by at least 24 hours to minimize observer memory bias. The mean lumen and outer wall surfaces for each carotid artery were then computed and matched on a point-by-point basis, from which the VWT was measured by taking the distance between each pair of corresponding points. 

\subsection{2D arc-length scaling (ALS) map}
With the 3D carotid outer wall surfaces and the VWT distributions superimposed on them, Chiu \textit{et al.} \cite{Chiu13} proposed the arc-length scaling (ALS) mapping method for flattening the surfaces onto a 2D non-convex L-shaped domain. Here we briefly describe the method and refer readers to the previous works by Chiu \textit{et al.} \cite{Chiu13b,Chiu16} for more details. The method begins with transforming the 3D carotid surface to a standard coordinate system, with the bifurcation (BF) located at the origin, the longitudinal direction of the common carotid artery (CCA) aligned with the z-axis, and the internal carotid artery (ICA) located at the upper half space of the coordinate system. The surface is then cut by two planes (Fig. \ref{fig:carotid_artery_flattening}\textbf{a}) and unfolded to a 2D L-shaped non-convex domain (Fig. \ref{fig:carotid_artery_flattening}\textbf{b}). The ICA and the CCA are respectively mapped to the top part and the bottom part of the L-shaped domain. The arc-length of transverse contours segmented from 2D transverse images resliced from 3D ultrasound images is rescaled so that the vertices on the carotid surface correspond to uniformly spaced grid points on the L-shaped domain. With the 2D L-shaped carotid template, a more systematic comparison between the VWT of different carotid arteries can be performed.

However, the ALS mapping approach does not consider any local geometric distortion produced by the 3D to 2D mapping. To reduce the geometric distortion, we consider deforming the ALS map and generate an area-preserving 2D carotid template.

\subsection{Area-preserving map via density-equalizing reference map (DERM)} 
Denote the 2D non-convex L-shaped domain obtained by the ALS mapping method by $D$. We would like to deform $D$ based on a prescribed density distribution. Let $\rho({\bf x},t)$ be the density at the location ${\bf x}$ on $D$ at time $t$. To achieve an area-preserving map, we set $\rho({\bf x},0)$ to be the area of each face of the carotid surface and consider a diffusion-based deformation. 

Here, we consider a more general version of the diffusion process when compared to the formulation \eqref{eqt:diffusion} described in Section \ref{sect:density_equalizing_map}. Consider the diffusion equation with diffusivity $\kappa$
\begin{equation}\label{eqt:derm_diffusion_kappa}
\frac{\p \rho}{\p t} = \nabla \cdot (\kappa \nabla \rho) = \kappa \Delta \rho + \nabla \kappa \cdot \nabla \rho.
\end{equation}
$\kappa$ is a differentiable function that we introduce for handling the non-convex corner of the L-shaped domain. To regularize the deformation around the non-convex corner, we slow down the diffusion process there by considering a differentiable function $\kappa$ with $\kappa \ll 1$ around the non-convex corner and $\kappa \approx 1$ at the regions far away from it. One possible choice of such $\kappa$ is 
\begin{equation} \label{eqt:kappa}
\kappa(x,y) = 1 - \left(1-\frac{1}{\sqrt{a}}\right) e^{-\frac{(x-p)^2+(y-q)^2}{\sqrt{a}}},
\end{equation}
where $(p, q)$ are the coordinates of the bifurcation point on $D$ and $a$ is the total area of $D$. Note that $\nabla \kappa$ in \eqref{eqt:derm_diffusion_kappa} can be expressed explicitly by taking partial derivatives on \eqref{eqt:kappa}. On the boundary edges of the L-shaped domain $D$, we enforce the no-flux boundary condition ${\bf n} \cdot \nabla \rho = 0$ where ${\bf n}$ is the unit outward normal on the boundary edges, so that the diffusion occurs only within $D$. This ensures that the subsequent diffusion-based deformation does not change the overall shape of the L-shaped domain.

With the diffusivity $\kappa$, the velocity field \eqref{eqt:velocity_field} becomes
\begin{equation}\label{eqt:derm_velocity_kappa}
{\bf v}({\bf x},t) = - \frac{\kappa \nabla \rho}{\rho}.
\end{equation}
Now, from the viewpoint of the reference map technique described in Section \ref{sect:reference_map}, we treat the L-shaped domain $D$ as a solid body and consider its deformation under the above velocity field ${\bf v}({\bf x},t)$, which is induced by the density gradient. We obtain the reference map ${\bm{\xi}}({\bf x},t)$ by solving the advection equation
\begin{equation} \label{eqt:derm_advection}
\frac{\p \bm{\xi}}{\p t}({\bf x},t) + {\bf v}({\bf x},t) \cdot \nabla \bm{\xi}({\bf x},t) = {\bf 0}.
\end{equation}
As $t \to \infty$, the density $\rho({\bf x},t)$ is equalized over $D$, and the associated reference map field $\bm{\xi}({\bf x}_{\text{final}},\infty)$ is a density-equalizing reference map. If the VWT at the location ${\bf X}$ on the initial 2D ALS map is described by a function $T({\bf X})$, then the VWT on the final area-preserving map will be given by $T(\bm{\xi}({\bf X},\infty))$. Equivalently, by considering the contour lines of constant $x$- and $y$- coordinates of $\bm{\xi}$, we obtain the final area-preserving map ${\bf x}_{\text{final}}$.


In the discrete case, suppose $D$ (including the top left empty space) is a rectangular grid consisting of $M\times N$ grid points, with grid spacing $h$ in both the $x$-direction and the $y$-direction. Here the top left empty space is included just for simplicity of the discretization and can be omitted in all the subsequent computations. Let the coordinates of the grid points be $(ih, jh)$, where $0 \leq i \leq M - 1$ and $0 \leq j \leq N-1$. The diffusion equation \eqref{eqt:derm_diffusion_kappa}, the velocity field \eqref{eqt:derm_velocity_kappa} and the advection equation \eqref{eqt:derm_advection} are discretized and then solved iteratively. Denote the step size by $\delta t$, and the density at the grid point $(ih, jh)$ at the $n$-th step by $\rho_{i,j}^n$. The discrete version of ${\bf v}$ and ${\bf X}$ is represented in a similar manner. 

Note that $\kappa$ and its derivatives $\kappa_x$, $\kappa_y$ can all be easily discretized. For the diffusion equation \eqref{eqt:derm_diffusion_kappa}, we use the central difference scheme to approximate $\nabla \rho$ and solve the equation by the implicit Euler method
\begin{equation} \label{eqt:derm_diffusion_kappa_discrete}
\resizebox{\hsize}{!}{$
\begin{split}
\frac{\rho_{i,j}^{n} - \rho_{i,j}^{n-1}}{\delta t} = &\kappa_{i,j}\frac{\rho_{i+1,j}^n+\rho_{i-1,j}^n+\rho_{i,j+1}^n+\rho_{i,j-1}^n-4\rho_{i,j}^n}{h^2} \\
&+ (\kappa_x)_{i,j} \frac{\rho_{i+1,j}^{n}-\rho_{i-1,j}^{n}}{2h} + (\kappa_y)_{i,j}\frac{\rho_{i,j+1}^{n}-\rho_{i,j-1}^{n}}{2h}.
\end{split}
$}
\end{equation}
The no-flux boundary condition for the diffusion equation is enforced using the following ghost node approach. At the four rectangular boundaries $(x,y) = (0,jh), ((M-1)h,jh), (ih,0), (0,(N-1)h)$ where $0 \leq i \leq M - 1$ and $0 \leq j \leq N-1$, and the two L-shaped boundaries $(x,y) = (p,jh), (ih, q)$ where $0 \leq ih \leq p$ and $0 \leq jh \leq q$, we suitably replace the terms $\rho_{i-1,j}^n$, $\rho_{i+1,j}^n$, $\rho_{i,j-1}^n$, $\rho_{i,j+1}^n$ on the right hand side in \eqref{eqt:derm_diffusion_kappa_discrete} by $\rho_{i,j}^n$ to ensure that there is no density flux orthogonal to the boundaries. Hence, the L-shaped domain will maintain its boundary shape throughout the density equalization process. If we represent $\rho^n$ as a column vector of size $MN \times 1$, \eqref{eqt:derm_diffusion_kappa_discrete} can be simplified as
\begin{equation} \label{eqt:derm_diffusion_kappa_discrete_matrix}
\rho^n = A^{-1} \rho^{n-1},
\end{equation}
where $A$ is an $MN \times MN$ matrix with $A = I - \delta t (\kappa \Delta+K_x+K_y)$, $K_x+K_y$ being the matrix representation of $\nabla \kappa \cdot \nabla$. Note that $A$ is a sparse matrix independent of $n$ and hence we only need to compute it once throughout the density-equalizing iterations. 

The discretization of the velocity field \eqref{eqt:derm_velocity_kappa} is achieved again using the central difference scheme
\begin{equation} \label{eqt:derm_velocity_discrete}
\left\{\begin{split}
({\bf v}_x)_{i,j}^n &= -\kappa_{i,j} \frac{\rho_{i+1,j}^{n}-\rho_{i-1,j}^{n}}{2h \rho_{i,j}^{n}},\\
({\bf v}_y)_{i,j}^n &= -\kappa_{i,j} \frac{\rho_{i,j+1}^{n}-\rho_{i,j-1}^{n}}{2h \rho_{i,j}^{n}}.
\end{split}\right.
\end{equation}

The advection equation \eqref{eqt:derm_advection} for updating the reference map $\bm{\xi}({\bf x},t)$ is solved using the upwind method:
\begin{equation} \label{eqt:derm_advection_discrete}
\resizebox{\hsize}{!}{$
\begin{split}
&\displaystyle \frac{\bm{\xi}_{i,j}^{n} - \bm{\xi}_{i,j}^{n-1}}{\delta t} \\
&= \left\{\begin{array}{ll}
- ({\bf v}_x)_{i,j}^{n} \frac{\bm{\xi}_{i,j}^{n-1} - \bm{\xi}_{i-1,j}^{n-1}}{h} - ({\bf v}_y)_{i,j}^{n}  \frac{\bm{\xi}_{i,j}^{n-1} - \bm{\xi}_{i,j-1}^{n-1}}{h} & \text{ if } ({\bf v}_x)_{i,j}^{n} > 0 \text{ and } ({\bf v}_y)_{i,j}^{n} > 0, \\
- ({\bf v}_x)_{i,j}^{n} \frac{\bm{\xi}_{i+1,j}^{n-1} - \bm{\xi}_{i,j}^{n-1}}{h} - ({\bf v}_y)_{i,j}^{n}  \frac{\bm{\xi}_{i,j}^{n-1} - \bm{\xi}_{i,j-1}^{n-1}}{h} & \text{ if } ({\bf v}_x)_{i,j}^{n} \leq 0 \text{ and } ({\bf v}_y)_{i,j}^{n} > 0,\\
- ({\bf v}_x)_{i,j}^{n} \frac{\bm{\xi}_{i,j}^{n-1} - \bm{\xi}_{i-1,j}^{n-1}}{h} - ({\bf v}_y)_{i,j}^{n}  \frac{\bm{\xi}_{i,j+1}^{n-1} - \bm{\xi}_{i,j}^{n-1}}{h} & \text{ if } ({\bf v}_x)_{i,j}^{n} > 0 \text{ and } ({\bf v}_y)_{i,j}^{n} \leq 0,\\
- ({\bf v}_x)_{i,j}^{n} \frac{\bm{\xi}_{i+1,j}^{n-1} - \bm{\xi}_{i,j}^{n-1}}{h} - ({\bf v}_y)_{i,j}^{n}  \frac{\bm{\xi}_{i,j+1}^{n-1} - \bm{\xi}_{i,j}^{n-1}}{h} & \text{ if } ({\bf v}_x)_{i,j}^{n} \leq 0 \text{ and } ({\bf v}_y)_{i,j}^{n} \leq 0.
\end{array}\right.
\end{split}
$}
\end{equation}

By iteratively solving \eqref{eqt:derm_diffusion_kappa_discrete_matrix}, \eqref{eqt:derm_velocity_discrete} and \eqref{eqt:derm_advection_discrete} until the density $\rho$ is fully equalized on the entire domain, we obtain the desired density-equalizing reference map $\bm{\xi}$. To obtain the associated area-preserving map ${\bf x}_{\text{final}}$, we denote $\bm{\xi} = (\bm{\xi}_1, \bm{\xi}_2)$. For every grid point $(ih,jh)$ on the initial 2D L-shaped domain obtained by the ALS mapping algorithm, where $i = 0,1,\cdots, M-1$ and $j = 0,1,\cdots,N-1$, the corresponding point of it in the final area-preserving map ${\bf x}_{\text{final}}$ is the intersection point of the contour lines $\bm{\xi}_1 = ih$ and $\bm{\xi}_2 = jh$. Note that in the discrete case, each contour line is represented as a piecewise linear curve. To check whether a line segment $\{(x_1^k, y_1^k), (x_1^{k+1}, y_1^{k+1})\}$ of a $\bm{\xi}_1$-contour intersects with a line segment $\{(x_2^l, y_2^l), (x_2^{l+1}, y_2^{l+1})\}$ of a $\bm{\xi}_2$-contour, it suffices to solve the following system of four linear equations in four unknowns $x^*, y^*, t_1, t_2$:
\begin{equation}
\left\{ \begin{array}{lll}
(x_1^{k+1}-x_1^k) t_1 &= &x^* - x_1^k,\\
(y_1^{k+1}-y_1^k) t_1 &= &y^* - y_1^k,\\
(x_2^{l+1}-x_2^l) t_2 &= &x^* - x_2^l,\\
(y_2^{l+1}-y_2^l) t_2 &= &y^* - y_2^l.
\end{array}\right.
\end{equation}
If $t_1, t_2 \in [0,1]$, then the two line segments intersect at $(x^*, y^*)$. Otherwise, they do not intersect. By tracking all the intersection points of the pairwise contour lines, the area-preserving map ${\bf x}_{\text{final}}$ can be obtained. 

As for the choice of the step size $\delta t$, note that the backward Euler scheme for the diffusion equation is unconditionally stable. Following the discussion in \cite{Choi18}, we perform a dimensional analysis on \eqref{eqt:diffusion} and see that an appropriate dimension of $\delta t$ is (length)$^2$. Also, as the density diffusion process is invariant under uniform rescaling of the input density $\rho$, $\delta t$ should be independent of the magnitude of $\rho$. Hence, it is desirable to have the step size in the form
\begin{equation} \label{eqt:step_size}
\delta t = \frac{\text{std}(\rho)}{\text{mean}(\rho)} \times a c,
\end{equation}
where $c$ is a dimensionless constant. The convergence criterion is 
\begin{equation} 
\frac{\text{sd}(\rho^n)}{\text{mean}(\rho^n)} \leq \epsilon,
\end{equation}
where $\epsilon$ is the error threshold. The algorithm is summarized in Algorithm \ref{alg:2d}. 
\begin{algorithm}[h]
\KwIn{A carotid surface $S$, the error threshold $\epsilon$, the maximum number of iterations allowed $n_{\max}$.}
\KwOut{An area-preserving map ${\bf x}_{\text{final}}$ on the 2D non-convex L-shaped domain.}
\BlankLine

Compute an initial map $f:S \to \mathbb{R}^2$ onto the 2D L-shaped domain using the ALS mapping \cite{Chiu13}\;

Compute the area $\rho$ of every face of $S$ and set $\rho$ as the density on the L-shaped domain\;

Set $\delta t = \frac{\text{std}(\rho)}{\text{mean}(\rho)} \times a c$\;

Compute $A = I - \delta t (\kappa \Delta+K_x+K_y)$ in \eqref{eqt:derm_diffusion_kappa_discrete_matrix} \;

Set $n = 0$ and $\rho^0 = \rho$\;

\Repeat{$\frac{\text{sd}(\rho^n)}{\text{mean}(\rho^n)} \leq \epsilon$ or $n \geq n_{\max}$}{ 

Solve $\rho^{n+1} = A^{-1} \rho^{n}$\;

Compute the velocity field ${\bf v}$ using \eqref{eqt:derm_velocity_discrete}\;

Update the reference map $\bm{\xi}$ using \eqref{eqt:derm_advection_discrete}\;

Update $n = n + 1$\;
}

By tracking the intersections of the contour lines $\bm{\xi}_1 = ih$ and $\bm{\xi}_2 = jh$ of the density-equalizing reference map $\bm{\xi} = (\bm{\xi}_1, \bm{\xi}_2)$ for all $i,j$, obtain the desired area-preserving map ${\bf x}_{\text{final}}$\;
\caption{Area-preserving carotid flattening via density-equalizing reference map (DERM)}
\label{alg:2d}
\end{algorithm}

Note that the original density-equalizing map (DEM) approach \cite{Gastner04} iteratively solves the diffusion equation \eqref{eqt:diffusion} with uniform diffusivity, obtains the velocity field \eqref{eqt:velocity_field} and tracks the displacement of every tracer by \eqref{eqt:displacement}. In the discrete case, updating each tracer by \eqref{eqt:displacement} requires an interpolation of the velocity field at its current location in every iteration. By contrast, the proposed DERM algorithm keeps track of the reference map field by \eqref{eqt:derm_advection}, which is fully Eulerian and hence no interpolation is needed throughout the iterations. 

\begin{figure*}[t!]
\centering
\includegraphics[width=0.73\textwidth]{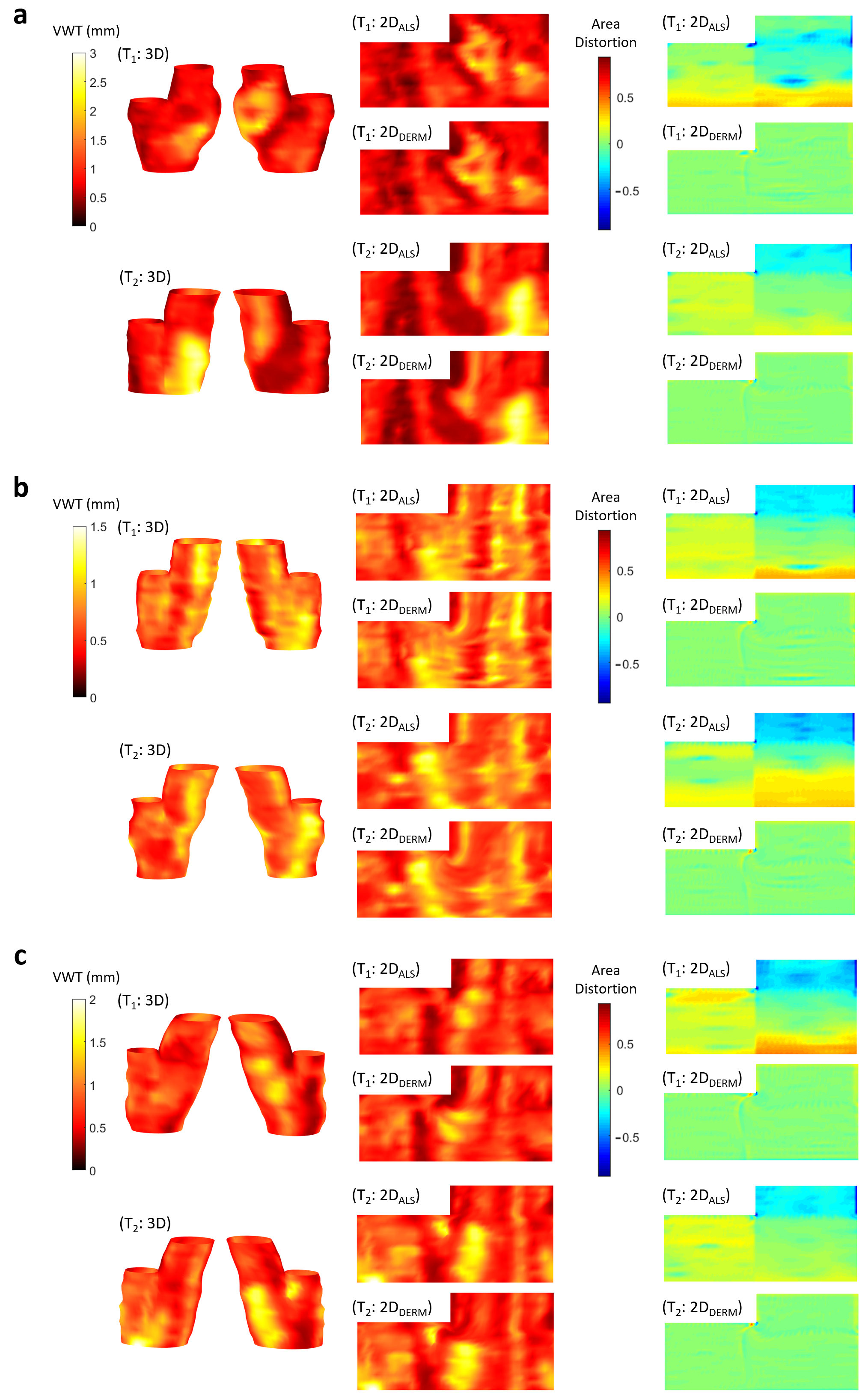}
\caption{Carotid flattening maps produced by the ALS mapping method \cite{Chiu13} and our area-preserving DERM method. \textbf{a}, \textbf{b}, \textbf{c} show the results for three subjects. For each subject, the carotid models constructed from the baseline and follow-up 3D ultrasound images are denoted by $T_1$ and $T_2$ respectively. The first column shows the 3D carotid surfaces color-coded and superimposed by their VWT distributions (with both front and back views). The second column shows the 2D VWT maps of the carotid surfaces produced by the ALS mapping method and our area-preserving DERM method. The third column shows the area distortion maps. }
\label{fig:artery_mapping_result}
\end{figure*}

\begin{figure*}[t!]
\centering
\includegraphics[width=0.9\textwidth]{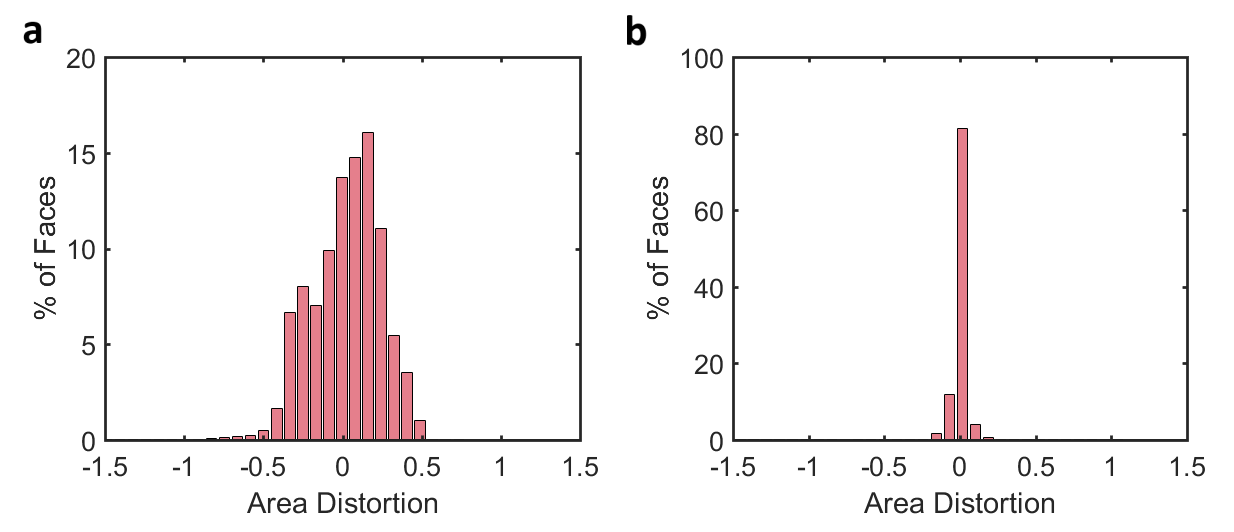}
\caption{Histograms of area distortions exhibited in 150360 quadrilateral faces (7518 per carotid model $\times$ 20 carotid models) in \textbf{a} the 2D ALS maps and \textbf{b} the 2D area-preserving maps produced by the proposed DERM algorithm.}
\label{fig:distortion_histograms}
\end{figure*}

\section{Results} \label{sect:experiment}
The 2D ALS mapping algorithm was implemented in C++ and took 7 seconds to process each arterial model. The proposed area-preserving DERM method was also implemented in C++ with OpenMP parallelization (with grid size $h = 1$, maximum number of iterations $n_{\max} = 500$, step size constant $c = 0.01$, error threshold $\epsilon = 10^{-3}$), taking 8 seconds on a PC with an Intel i7-6700K quad-core processor and 16GB RAM for each arterial model. The sparse linear systems were solved using the biconjugate gradient stabilized method (\texttt{BiCGSTAB}) in the C++ library Eigen. The visualization and statistics were produced using MATLAB.

The first column of Fig. \ref{fig:artery_mapping_result} shows the front and back views of six example carotid models from three subjects with their VWT distributions color-coded and superimposed. For each subject, the carotid models at baseline and follow-up are labeled as $T_1$ and $T_2$ respectively. We flattened the 3D models onto the 2D L-shaped domain using the ALS mapping method \cite{Chiu13} and our area-preserving DERM method (see the middle column of Fig. \ref{fig:artery_mapping_result}). Each of the 2D ALS and area-preserving flattening maps was made up of 7518 quadrilateral elements. The area distortion associated with each quadrilateral face was quantified by the logged area distortion ratio defined below: 
\begin{equation}
d = \log \frac{\text{Area of the face on the flattened map}}{\text{Area of the face on the 3D carotid model}}.
\end{equation}
A perfectly area-preserving map will result in $d = 0$ for all faces. A positive $d$ at a local region indicates that the region is enlarged on the L-shaped 2D carotid template, while a negative $d$ indicates that the region is shrunken. $d$ represents enlargement and shrinkage of the same factor in an equal magnitude. For example, a face with the area halved and another with the area doubled after being flattened have $d = -0.693$ and $0.693$ respectively. With this symmetric property, the logged ratio $d$ is easier to interpret than the linear ratio. The last column of Fig. \ref{fig:artery_mapping_result} shows the flattened maps with $d$ color-coded and superimposed. In these examples, the area distortion exhibited in the ALS map was largely corrected by the proposed DERM algorithm. We further evaluated the area distortion of the 150360 quadrilateral elements in all the 20 carotid models (7518 per carotid model $\times$ 20 models) before and after the application of the proposed algorithm. Fig.~\ref{fig:distortion_histograms} shows the area distortion histograms, from which it can be observed that the area distortion obtained by the proposed method is much more concentrated at 0 when compared to that by the ALS mapping method. 
 
\begin{figure*}[t!]
\centering
\includegraphics[width=0.8\textwidth]{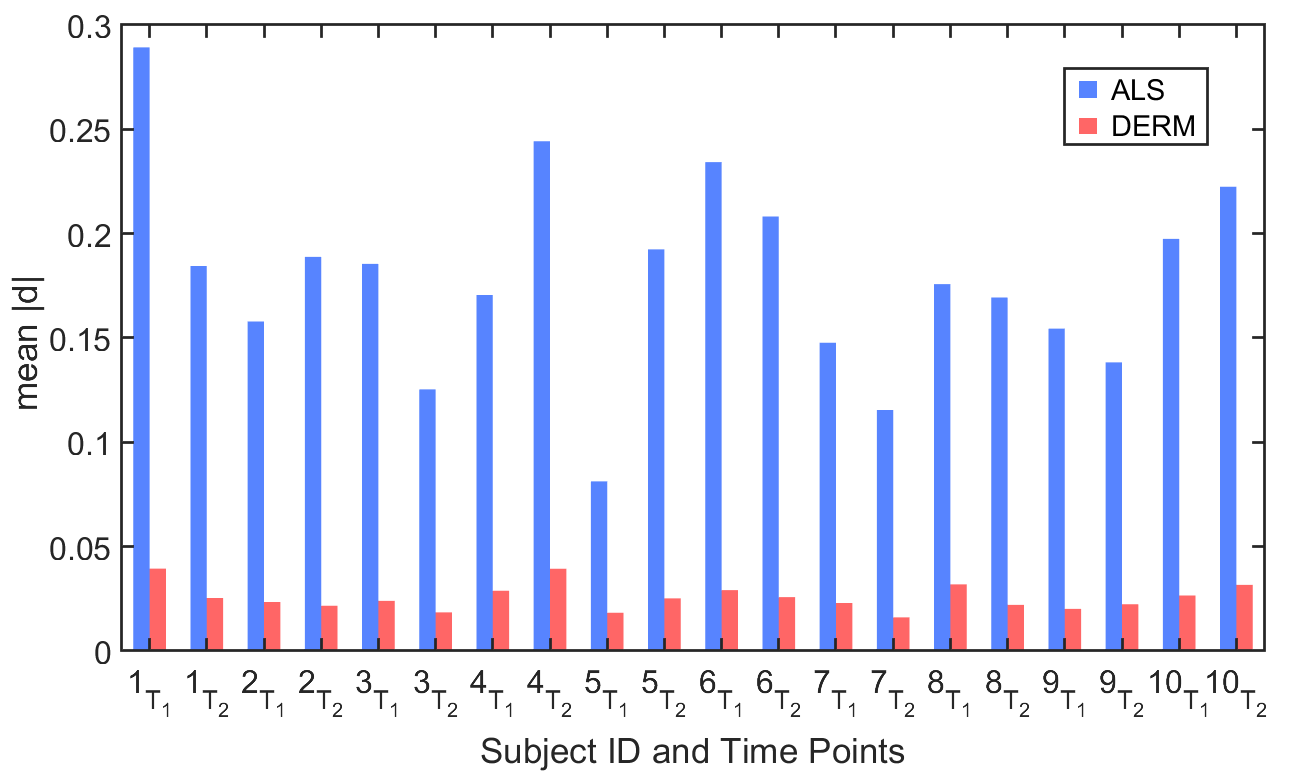}
\caption{Average absolute area distortion $\text{mean}(|d|)$ of 20 carotid models. The carotid models constructed from the baseline and follow-up 3D ultrasound images for each subject are denoted by the subscripts $T_1$ and $T_2$ respectively.}
\label{fig:distortion_individual}
\end{figure*}

We also quantified the performance of the proposed flattening algorithm for each of the 20 carotid models by the average value of $|d|$ over the corresponding 2D maps. Fig.~\ref{fig:distortion_individual} shows the values $\text{mean}(|d|)$ of the 2D ALS map and the 2D area-preserving map produced by the proposed DERM method for each carotid model. The proposed method attained a reduction in area distortion by over 80\% on average. A two-sample $t$-test shows that the average area distortion is significantly reduced by the proposed method ($P < 10^{-6}$). 

\begin{figure*}[t]
\centering
\includegraphics[width=0.9\textwidth]{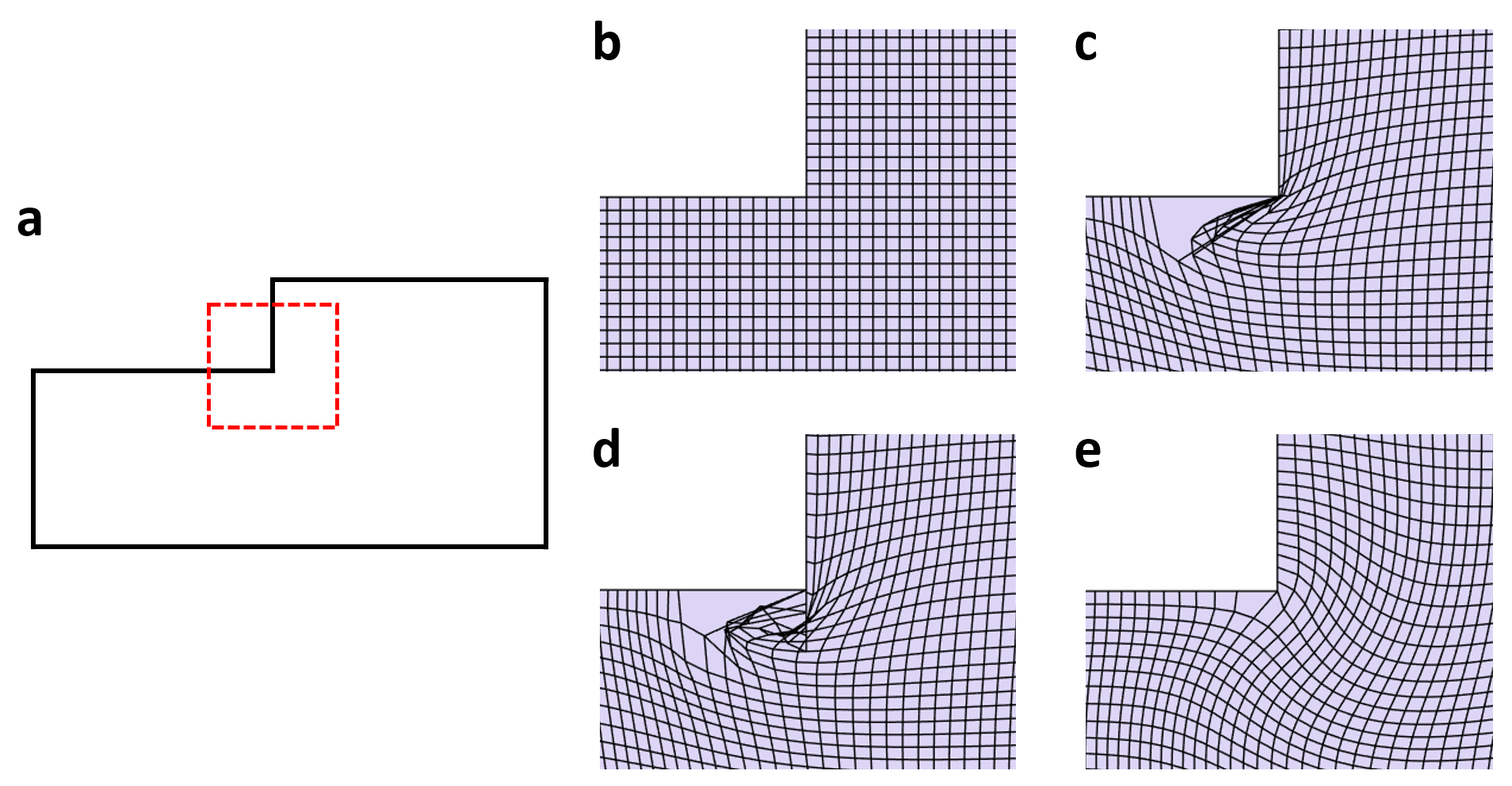}
\caption{Comparison between various mapping methods for handling the non-convex corner of the L-shaped domain. We consider zooming into the non-convex corner of the L-shaped domain as illustrated in \textbf{a}. In \textbf{b}, \textbf{c}, \textbf{d}, \textbf{e}, the mapping results at that region produced by the ALS mapping method \cite{Chiu13}, the density-equalizing map (DEM) \cite{Gastner04}, the shape-prescribed density-equalizing map (SPDEM) \cite{Choi18} and the proposed DERM method are respectively shown. It can be observed that our method is advantageous in adjusting for the geometric variability of the carotid surfaces without causing any overlaps.}
\label{fig:comparison_corner}
\end{figure*}

In addition to minimizing area distortion, the proposed method contributed in the elimination of overlapping cells when mapping to a non-convex domain. Fig. \ref{fig:comparison_corner} compares the performance of three methods aiming to correct the area distortion associated with ALS mapping. Figs. \ref{fig:comparison_corner}\textbf{c}-\textbf{e} show the results generated at the non-convex corner of the L-shaped domain by the density-equalizing map (DEM) \cite{Gastner04}, the shape-prescribed density-equalizing map (SPDEM) \cite{Choi18} and the proposed DERM method. For both DEM and SPDEM, the density was set to be the face area for computing area-preserving flattening maps. The free boundary condition in DEM was replaced by the no-flux boundary condition for enforcing the L-shape. It can be observed that both DEM and SPDEM produce undesirable overlaps around the non-convex corner of the L-shaped domain (Fig. \ref{fig:comparison_corner}\textbf{c}, \textbf{d}) due to the large density gradient (i.e. face area difference) between the ICA and CCA generated by ALS mapping as shown in the last column of Fig. \ref{fig:artery_mapping_result}. Since the proposed DERM method has a non-constant diffusion coefficient $\kappa$ in the diffusion process and involves the reference map technique for producing a smoother deformation, we are able to obtain an area-preserving flattening map without any overlaps (Fig. \ref{fig:comparison_corner}\textbf{e}), thereby guaranteeing bijectivity. We quantified the above observation by evaluating the total overlapping area in each flattened map, computed by taking the difference between the sum of area of all quadrilateral faces in the flattened map and the area enclosed by the boundary of the L-shaped domain. This overlap metric was computed for the entire set of 20 arteries. The mean and the standard deviation for DEM and SPDEM were 23.3 $\pm$ 13.9 and 30.5 $\pm$ 23.3 respectively. Since the proposed algorithm guarantees bijectivity, this overlap metric is 0 for all arterial models.

\section{Discussion and Conclusion} \label{sect:conclusion}
In this work, we have proposed a novel method for producing area-preserving flattening maps for visualization and consistent quantification of the VWT distributions on carotid surfaces. Our method first takes the 2D L-shaped non-convex domain produced by the ALS mapping algorithm \cite{Chiu13} as an initial map. Then, we improved the density diffusion process \cite{Choi18,Gastner04} and combined it with the reference map technique \cite{Kamrin12,Valkov15}. Density-equalizing maps \cite{Gastner04} have been widely used in sociological data visualization and was recently introduced for the computation of area-preserving surface mapping \cite{Choi18}. However, the bijectivity of the maps obtained by these methods is only guaranteed when the maps are with free boundary constraints or with a convex target shape. To overcome this limitation so as to produce area-preserving maps with the non-convex L-shape, we extended the formulation of density diffusion in these methods to allow for a non-constant diffusivity that effectively regularizes the density gradient around the non-convex corner.

Another limitation of the previous methods is that the deformation of each node on the domain was tracked  individually without considering the neighboring spatial information, which results in an unsmooth mapping result under a large deformation. The reference map technique \cite{Kamrin12,Valkov15} was originally developed for tracking large physical deformations of a solid body smoothly. For the first time in the world, the reference map technique was integrated with the density-equalization process to generate a smooth and area-preserving flattened map. Although this technique was applied on carotid mapping, it can be applied to facilitate the interpretation of spatial distributions on the surfaces of other organs, such as brain ventricles and kidneys \cite{Davies02}. Carotid mapping described in this paper is particularly challenging due to the non-convex nature of the carotid template. A highlight of the proposed algorithm is that it allows bijective (one-to-one) mapping even at the non-convex corner at the carotid bifurcation, whereas previous density-equalizing approaches do not guarantee bijectivity in a non-convex domain (Fig. \ref{fig:comparison_corner}). 

The proposed algorithm shares with previous area-preserving algorithms \cite{Zhu05,Chiu08} that the area distortion could not be completely eliminated, although the proposed method reduced the area distortion in a statistically significant manner. This limitation can be explained by the finite grid size and step size used in our numerical scheme. A possible way to further reduce the area distortion without slowing down the computation is to extend our algorithm with an adaptive framework, which will allow us to take adaptive time steps to handle specific regions with relatively large distortion. This strategy will be implemented and our hypothesis will be validated in a future investigation. 

Although the density-equalizing approach introduced here was used to generate area-preserving carotid maps, its application can be easily extended to non-rigid carotid registration (i.e. finding an optimal deformation to match carotid images). 3D ultrasound images are acquired at baseline and a follow-up imaging session in serial monitoring of the carotid disease. Even if no physiological changes are expected, such as in the case where patients are presented with stable disease and the time interval between the baseline and follow-up imaging session is as short as two weeks \cite{Egger08}, there would be variance in the vessel-wall-plus-plaque thickness (VWT) maps due to different patient position and neck orientation that could only be corrected by non-rigid registration techniques. The proposed pipeline can be used to register the VWT maps obtained in the two imaging sessions in a non-rigid manner. The improved reproducibility of the VWT maps thus produced may play a role in increasing the sensitivity of biomarkers that quantify the change in VWT distribution for evaluation of new therapies. One example of such biomarker is the mutual-information-weighted biomarker used to quantify the weighted VWT-Change average exhibited in a one-year longitudinal study involving patients who received placebo and Vitamin B tablets \cite{Cheng17}. Although a statistically significant difference was found between the two treatment groups, thereby establishing the effect of Vitamin-B treatment on atherosclerosis, only rigid registration was performed to align the arteries imaged at baseline and the follow-up session before the construction of VWT maps. The non-rigid registration capability of the proposed algorithm has the potential to further increase the sensitivity of the biomarkers characterizing VWT-Change. Such an improvement would reduce the sample size required to establish effects of new medical treatments on atherosclerosis, thereby increasing the cost-effectiveness of clinical trials.

\end{document}